\documentclass[a4paper,fleqn,usenatbib]{mnras}

\usepackage{mathptmx}
\usepackage[T1]{fontenc}
\usepackage{ae,aecompl}
\usepackage{ulem}

\usepackage{graphicx}	
\usepackage{amsmath}	
\usepackage{amssymb}	
\usepackage{txfonts}

\title[High-confidence radio QPO in the BL Lac PKS J2134-0153
]{Detection of a possible high-confidence radio quasi-periodic oscillation in the BL Lac PKS J2134-0153
}

\author[G.-W. Ren et al.]{
Guo-Wei Ren,$^{1}$
Nan Ding,$^{2}$\thanks{E-mail: orient.dn@foxmail.com (N, D)}
Xiong Zhang,$^{3}$\thanks{E-mail: ynzx@yeah.net (X, Z)}
Rui Xue,$^{4}$
Hao-Jing Zhang,$^{3}$
\newauthor
{}Ding-Rong Xiong,$^{5}$
Fu-Ting Li,$^{3}$
Hui Li,$^{2}$
\\
$^{1}$Department of Astronomy, Xiamen University, Xiamen 361005,P. R. China\\
$^{2}$School of Physical Science and Technology, Kunming University, Kunming 650214,P. R. China\\
$^{3}$College of Physics and Electronics, Yunnan Normal University, Kunming 650500,P. R. China\\
$^{4}$College of Physics and Electronic Information Engineering, Zhejiang Normal University, Jinhua 321004,P. R. China\\
$^{5}$Yunnan Observatories, Chinese Academy of Sciences, 396 Yangfangwang, Guandu District, Kunming, 650216,P. R. China\\
}

\date{Accepted XXX. Received YYY; in original form ZZZ}

\pubyear{2015}

\begin{document}
\label{firstpage}
\pagerange{\pageref{firstpage}--\pageref{lastpage}}
\maketitle

\begin{abstract}
We have searched quasi-periodic oscillations (QPOs) for BL Lac PKS J2134-0153 in the 15 GHz radio light curve announced by the Owens Valley Radio Observatory 40-m telescope during the period from 2008-01-05 to 2019-05-18, utilizing the Lomb-Scargle periodogram (LSP) and the weighted wavelet Z-transform (WWZ) techniques. This is the first time that to search for periodic radio signal in BL Lac PKS J2134-0153 by these two methods. These two methods consistently reveal a QPO of $4.69 \pm 0.14$ years ($> 5\sigma$ confidence level). We discuss possible causes for this QPO, and we expected that the binary black holes scenario, where the QPO is caused by the precession of the binary black holes, is the most likely explanation. BL Lac PKS J2134-0153 thus could be a good binary black hole candidate. In the binary black holes scenario, the distance between the primary black hole and the secondary black hole $a\sim1.83\times 10^{16}$ cm.

\end{abstract}




\section{Introduction}

Blazars, including flat spectrum radio quasar (FSRQ) and BL Lacertae object (BL Lac), is one of the most special subclasses of AGNs. Because their relativistic jets point to the observers, they emit bright light in a non-thermal form, cover a wide range of the electromagnetic spectrum, from radio to high-energy gamma-rays (Urry \& Padovani. 1995; Padovani. 2017). For the same reason, blazars show rapid and dramatic changes in the entire electromagnetic spectrum, display high and variable polarizations at radio and optical bands (e.g., Ulrich et al. 1997; Gupta et al. 2018; Ding et al. 2019).

Blazar light curves display a variety of features frequently, studies of the features enable us to explore their internal structures, physical properties, and radiation processes, e.g. Urry et al. (1995); Ulrich et al. (1997). Some of these studies reveal a particularly intriguing phenomenon is quasi-periodic oscillations (QPOs). The QPOs have been found in the multi-frequency blazar light curves including radio, optical, X-ray and $\gamma$-ray, the timescales of the reported periodicity between a few hours and a few years (Bhatta et al. 2016,2017; Zola et al. 2016).

In the radio band, quasi-periodic signals with different timescales are found in different blazars: Raiteri et al. (2001) reported QPOs with periods between about one year and several years in the BL Lac AO 0235. Bhatta et al. (2017) found a strong signal of QPO with 270 days period in the BL Lac  PKS 0219-164. King et al. (2013) found persistent $\sim $150-day periodic modulation in the 15 GHz observations in the FSRQ J1359+ 4011. Li et al. (2017) found an approximately 370-day quasi-period in the BL Lac 1ES 1959+650, however, this candidate QPO isn't statistically significant, and it can be attributed by the stochastic red noise process. Jaron et al. (2017) determined periodicities of $\sim$ 15 h in LS I $+61 ^{\circ}303$. Koay et al. (2019) identified 20 sources exhibiting significant flux density variations on 4-day timescales from the sample of 1158 radio-selected blazars span by the Owens Valley Radio Observatory (OVRO) 40-m telescope, and the origin of the days' time-scale variability maybe is interstellar scintillation. Search for the QPO of the radio  band is one of the methods to find a good binary black hole candidate.

In this paper, we analyze the long term ($\sim$11.4 years) observations of BL Lac PKS J2134-0153 and report our discovery of a high confidence QPO (> 5$\sigma$ confidence level) of the radio flux variability. In section 2, we introduce two long-period variability analysis methods of the blazar including the Lomb-Scargle Periodogram(LSP) and the Weighted Wavelet Z-transform (WWZ) techniques. In section 3, we employ these two methods to analyze the quasi-periodic behavior of PKS J2134-0153 and estimate the confidence of the QPO; In section 4, we discuss three scenarios to explain the QPO behavior, and estimate the distance between the primary black hole and the secondary black hole; Our conclusions be summarized in section 5.

\section{Long-period variability analysis methods of blazar}

The most direct means of estimating the time series is used by the Fourier transform, the frequency domain analysis method. Brigham et al. (1988) proposed Fast Fourier Transform  makes the Fourier transform occupy an important position in the field of discrete signal processing. The discrete Fourier transform shifts the signal from the time domain to the frequency domain for analysis. By observing the power spectrum of the discrete signal, it is easy to determine the phase and amplitude of the discrete signal with a certain characteristic frequency, thus revealing the frequency domain characteristics of the signal.

However, the traditional discrete Fourier transform in the above equation is no longer part applicable for non-uniformly sampled signals. The cause is that the Discrete Fourier Transform (DFT) is very sensitive to the sampling interval. It adds directly to the DFT to process the non-uniformly sampled signal. It is easy to have false periodic components in the spectrum ( Foster et al. 1996). Therefore, various improved algorithms based on discrete Fourier transform have emerged, which addresses the problem caused by non-uniform sampling to some extent. The following describes several algorithms of non-uniform sampling signal period extraction based on frequency domain analysis (Fourier transform).

\subsection{Lomb-Scargle Periodogram}
\label{sec:maths} 

LSP is a DFT-based periodic extraction algorithm proposed and improved by Lomb and Scargle. It solved the non-periodic signal generated by the non-uniform sampling interval to some extent(Scargle et al. 1981,1982; Lomb et al. 1976), avoided the interpolation of non-uniform sampling time series, and considered the effect of non-uniform sampling on amplitude and phase. For the non-uniform sampling time series $x\left ( t_{i} \right )$, $i=1,2,3\cdots , N$, the power spectrum is defined as:

\begin{equation}
\begin{aligned}
P_{LS}\left ( f \right )=&\frac{1}{2}\times \left [ \frac{\left \{ \sum_{i=1}^{N}x\left ( t_{i} \right )cos\left [ 2\pi f\left ( t_{i}-\tau  \right ) \right ] \right \}^{2}}{\sum_{i=1}^{N}cos^{2}\left [ 2\pi f\left ( t_{i}-\tau  \right ) \right ]}\right. \\
& \left.+\frac{\left \{ \sum_{i=1}^{N}x\left ( t_{i} \right )sin\left [ 2\pi f\left ( t_{i}-\tau  \right ) \right ] \right \}^{2}}{\sum_{i=1}^{N}sin^{2}\left [ 2\pi f\left ( t_{i}-\tau  \right ) \right ]}\right ]^{2}
\end{aligned}
\end{equation}
Where $f$ is the test frequency, and $\tau$ is the time offset, which can be obtained by the following formula.

\begin{equation}
tan\left ( 2\pi f\tau  \right )=\frac{\sum_{i=1}^{N}sin2\pi ft_{i}}{\sum_{i=1}^{N}cos2\pi ft_{i}}
\end{equation}

\subsection{Weighted Wavelet Z-transform}
\label{sec:maths} 

The astronomical observation signals are affected by the observation season, the weather, and the phase of the moon. The observation data are often non-equally spaced, and some intervals are large, even the intervals appear periodically, it has brought great difficulties to our analysis. In practice, when using Fourier transform or wavelet transform to process non-equal interval data, the interpolation method is used, but this has a great influence on the authenticity of the data. A pseudo period occurs in the transform spectrum when non-equal interval data is handled by Fourier transform.

Foster et al. (1996a) proposed the idea of vector projection for the wavelet transform to process non-equal interval data. He pointed out that if the wavelet transform is regarded as the projection of the vector, the analysis result can be significantly improved, and the period can be obtained more accurately, the stability of the period can revealed.

We mainly followed the WWZ procedure described in Foster et al. (1996b), it defined the WWZ as follows according to his proposed Z statistic:
\begin{equation}
Z=\frac{\left ( N_{eff} -3\right )V_{y}}{2\left ( V_{x} -V_{y}\right )}
\end{equation}

Where $N_{eff}=\frac{\left ( \sum w_{\alpha } \right )^{2}}{\sum w{_{\alpha }}^{2}}$ is the number of valid data, $V_{x}=\frac{\sum _{\alpha }w_{\alpha }y^{2}\left ( t_{\alpha } \right )}{\sum _{\beta }w_{\beta }}-\left [ \frac{\sum _{\alpha }w_{\alpha }y^{2}\left ( t_{\alpha } \right )}{\sum _{\beta }w_{\beta }} \right ]^{2}$ is the weighted deviation of the model function.

\section{The variability analysis and results of PKS J2134-0153}

In this paper, we used the WWZ and LSP methods to analyze the 15 GHz radio observations of the source PKS J2134-0153 announced by OVRO, we found that the variability period is about 4.69 years.

\subsection{Periodicity Search}

OVRO announced 15 GHz radio observation data from a source BL Lac PKS J2134-0153 (Z=1.285,(e.g. Truebenbach \& Darling. 2017)) observed with the 40-meter telescope from January 5, 2008, to May 18, 2019, for a total of $4152$ days ($\sim$ 11.4 years), a total of 638 data points, as can be seen from fig 1, there is a clear outbreak activity in this source, and three distinct peaks can be clearly seen, from the light curve, we can visually see that this source has a QPO between 1600 days and 1800 days. We proceeded a preliminary analysis of these data, in the 15 GHz radio band data of this source, the minimum flux is 1.63 Jy, the maximum flux is 2.84 Jy, the average flux is 2.22 Jy, and the standard deviation is 0.31. To quantify the observed variability, we estimate the variability amplitude (VA), which represents the peak-to-peak oscillation, and the fractional variability (FV), which represents the mean variability. We estimate the amplitude of the peak-to-peak variation using the relationship given in (Heidt \& Wagner. 1996):

\begin{figure*}
    \begin{center}
	\includegraphics[width=7in]{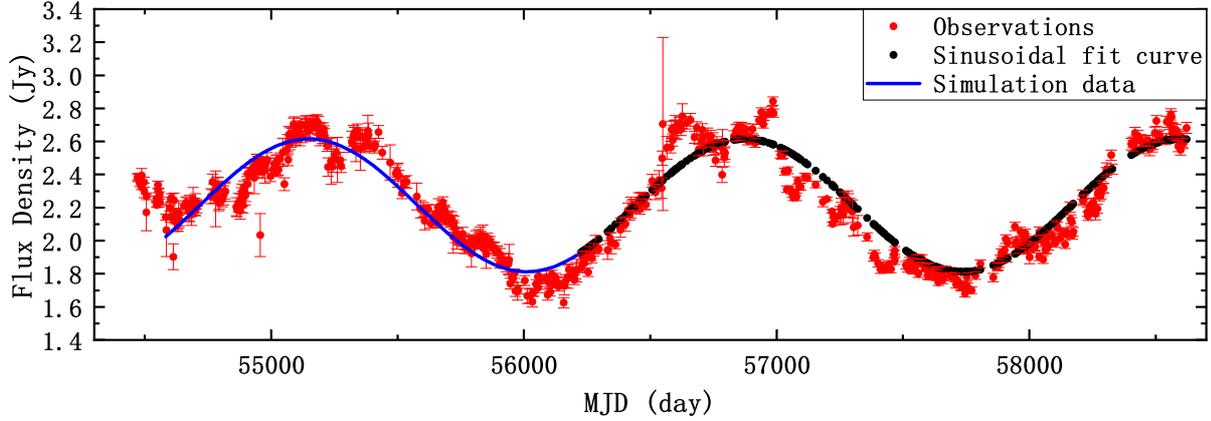}
    \end{center}
    \caption{15 GHz radio original light curve of BL Lac PKS J2134-0153 from January 5, 2008 to May 18, 2019 (red solid dots), the sinusoidal fitting curve for the first half of observations, and the predictions(black solid dots) for the second half of observations based on the fit of the first half of observations.}
    \label{fig:figure8}
\end{figure*}

\begin{equation}
VA=\sqrt{\left ( A_{max}-A_{min} \right )^{2}-2\sigma ^{2}}
\end{equation}
Where $A_{max}$, $A_{min}$, $\sigma$ represent the maximum value, minimum value, and average value of the magnitude errors in the radio observation data, respectively. Similarly, we can calculate the fractional variance FV with a mean flux of $\left \langle F \right \rangle$ with $S^{2}$ variance, and $\left \langle \sigma _{err}^{2} \right \rangle$ represent mean squared uncertainties, as stated in the formula given in (Vaughan et al. 2003):

\begin{equation}
F_{var}=\sqrt{\frac{S^{2}-\left \langle \sigma _{err}^{2} \right \rangle}{\left \langle F \right \rangle^{2}}}
\end{equation}

The error of fractional variability can be estimated as stated in the formula given in (Aleksi{\'c et al. 2015):

\begin{equation}
\sigma _{F_{var}}=\sqrt{F{_{var}}^{2}+\sqrt{\frac{2}{N}\frac{\left \langle e{_{err}}^{2} \right \rangle^{2}}{\left \langle F \right \rangle^{4}}+\frac{4}{N}\frac{\left \langle \sigma {_{err}}^{2} \right \rangle}{\left \langle F \right \rangle^{2}}F{_{var}}^{2}}}-F_{var}
\end{equation}

In this case, we can get $VA=1.21$, $F_{var}=0.14\pm 0.014$ through the above formula, which shows that there is a modest change in this source during this time.

Fig 1 shows the original light curve of PKS J2134-0153, we divide these 638 data evenly into two parts, i.e. each part contains 319 data points, and the blue curve is a sinusoidal fitting curve, which is plotted to help to indicate the modulation in the first half of observations, then we made predictions (black solid dots for the second half of the observations based on the fit of the first half of observations. Then, in order to analyze the correction between the simulated data and the original observations, we performed Pearson test. The result is shown in fig 2. The Person test gives a coefficient of correlation $R=0.950$ and a significance level $P< 0.0001$, the best liner fitting equation given by unary linear regression as $y=\left ( 0.81\pm 0.14 \right )x+\left ( 0.44\pm 0.03 \right )$, and $k=0.81\pm 0.14$. The results show that there are a positive correlation between the simulated data and the real observations.

\begin{figure}
	\includegraphics[width=9cm,height=6.1cm]{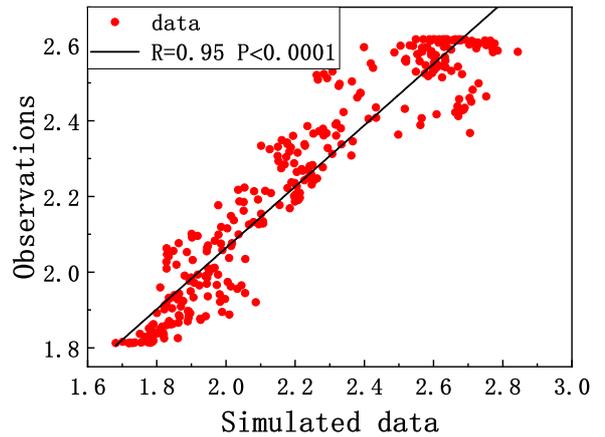}
    \caption{The correlation between the simulated data and the original observations. The black solid line is the best linear fitting $\left ( p < 0.0001 \right )$.}
    \label{fig:figure11}
\end{figure}

When analyzing observations by traditional periodic analysis, the observation data is required to be continuously uniform in the time domain. However, due to the influence of actual factors such as weather and observation equipment, the actual observation data cannot meet the requirements for the traditional method. Therefore, the observation data need to be interpolated or zero-filled. These processes have a significant impact on the results of the periodic analysis.

In order to search for QPOs of the source PKS J2134-0153, we first analyzed the 15 GHz flux density variations of the LSP method, which is one of the most common methods of time series analysis. Considering that the longest time interval between two observations of PKS J2134-0153 was 55 days. We averaged all flux density measurements over 60-day bins to smooth the short-term variability (Ulrich et al. 1997). The LSP of J2134-0153 light curve was computed for the minimum and maximum frequencies of $f_{min}=1/4152$ d, and $f_{max}=1/120$ d, respectively. What needs to be emphasized is that the estimate of the total number of periodogram frequencies, $n_{0}$, is critical to the evaluation of the periodogram. In this work, we evaluate the total number of periodogram frequencies using

\begin{equation}
N_{eval}=n_{0}Tf_{max}
\end{equation}
Where $n_{0}=10$, and $T=4152$ days represents the spanning time of the observations (Ulrich et al. 1997).

We analyzed the $\sim 11.4$ years long OVRO light curve of PKS J2134-0153, the result showing a distinct peak stands out around the time scale of $P_{obs}=1743\pm 401$ days, take the half-width at the half-maximum (HWHM) of the peak for the uncertainty in the value of QPO period.

To further confirm the presence of the above QPO by the different method, we performed the WWZ analysis of the entire light curve, in order to reduce the amount of calculation, we also averaged all flux density measurements over 60 days, perform WWZ transform on the processed data, the WWZ transform of PKS J2134-0153 light curve was computed for the minimum and maximum frequencies of $f_{min}=1/4152$ d, and $f_{max}=1/120$ d, respectively. Fig 5 gives the result, showing a distinct peak stands out around the time scale of $1672\pm 332$ days.

\begin{figure}
	\includegraphics[width=9cm,height=6.1cm]{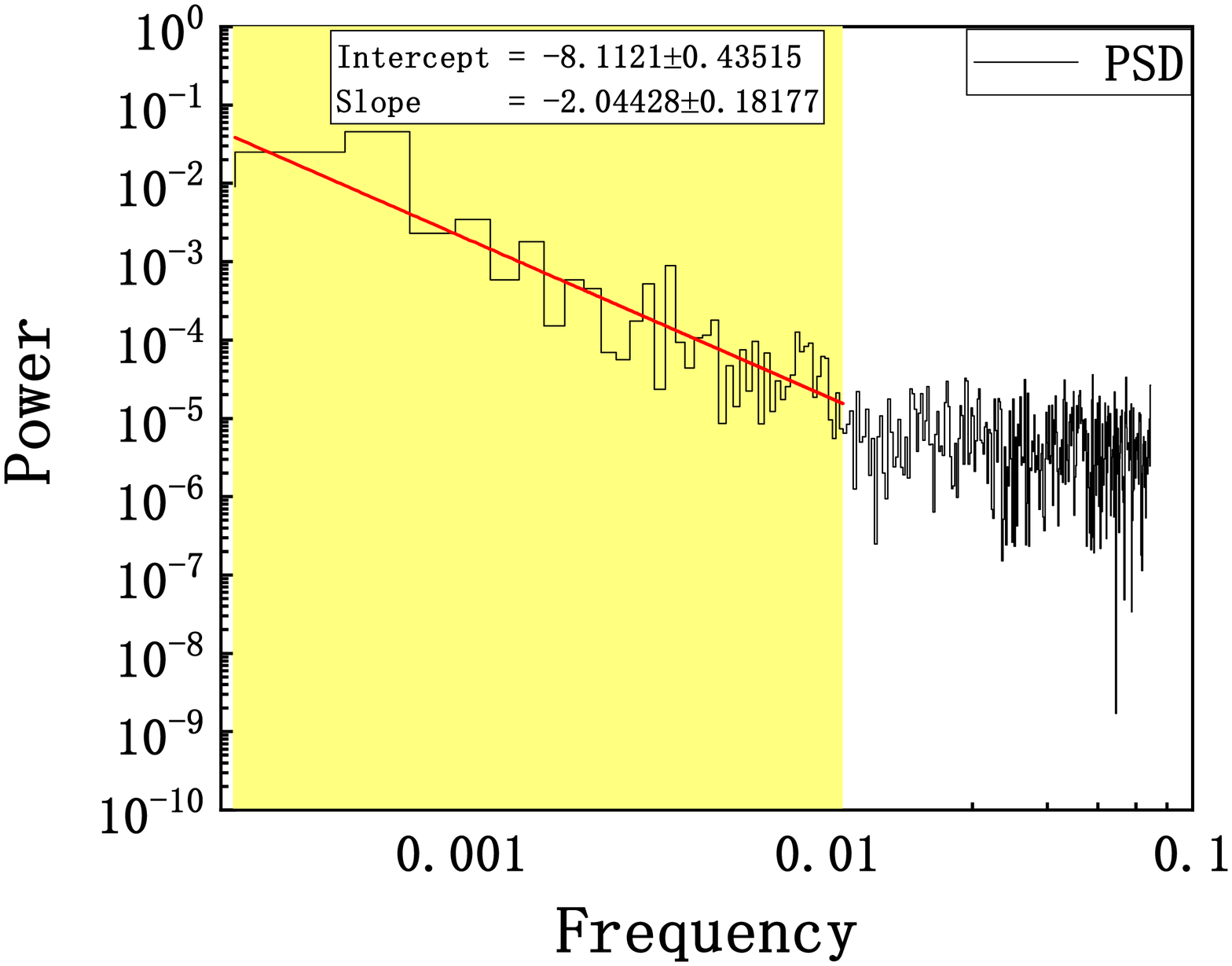}
    \caption{The PSD of the actual observations in BL Lac PKS J2134-0153, we fitted the part where frequency $<$ 0.01 and found the slope is -2.04428 ( $\sim$ -2.0 ).}
    \label{fig:f1}
\end{figure}

\begin{figure}
	\includegraphics[width=9cm,height=6.1cm]{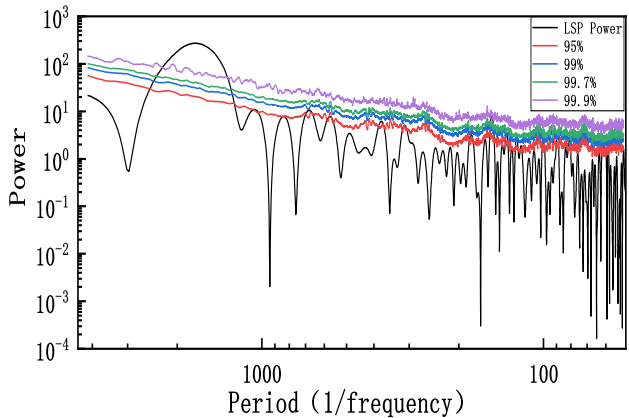}
    \caption{Lomb-Scargle periodogram of the $\sim $ 11.4 years long 15 GHz observations of the blazar BL Lac PKS J2134-0153. Simulate 20000 light curves by the Monte Carlo method described in Timmer \& Koenig. 1995, once the 20000 light curves were simulated by using even sampling interval, and their LSP was computed. At last, using the spectral distribution of the simulated light curves, local $95\%$, $99\%$, $99.7\%$, and $99.9\%$ confidence contour lines were evaluated,. The  red, blue, green, and purple solid lines represent the confidence level of $95\%$, $99\%$, $99.7\%$, and $99.9\%$, respectively.}
    \label{fig:figure9}
\end{figure}

\begin{figure}
	\includegraphics[width=9cm,height=6.1cm]{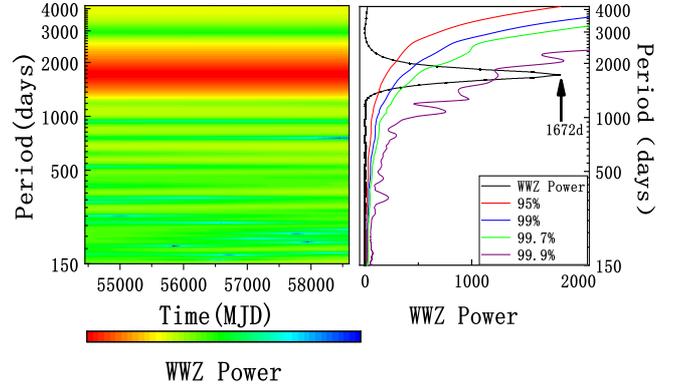}
    \caption{Weighted wavelet z-transform of the light curve presented. The left panel shows the distribution of color-scaled WWZ power (with red most intense and blue lowest) in the time-period plane; the right panel shows the time-averaged WWZ power (solid black curve) as a function of period showing a distinct peak stands out around the time scale of $1672\pm 332$ days; Simulate 20000 light curves by the Monte Carlo method described in Timmer \& Koenig. 1995, once 20000 light curves were simulated by using even sampling interval, and their WWZ power was computed. At last, utilizing the spectral distribution of the simulated light curves, local $95\%$, $99\%$, $99.7\%$, and $99.9\%$ confidence contour lines were evaluated. The  red, blue, green, and purple solid lines represent the confidence level of $95\%$, $99\%$, $99.7\%$, and $99.9\%$, respectively.}
    \label{fig:figure10}
\end{figure}

\subsection{Significance Estimation}

We detect periodicity in the light curve by LSP and WWZ methods respectively, and the light curves of blazars show the evidence of correlated noise (e.g., Press. 1978; Abdo, et al. 2010; Covino, Sandrinelli \& Treves. 2019). The periodic variability of blazars may be produced by the red noise process. In this case, this issue was solved by the power response method, which is usually used to characterise the power spectrum density of AGN. The periodogram of the source was modeled by a power-low Power spectral density (PSD): $P\propto f^{-\alpha }+C$; $f$ represent the frequency, $\alpha$ represent the slope of spectral, and $C$ represents the Poisson noise level(e.g., Bhatta. 2018; Nilsson, et al. 2018).We have calculated the PSD of the actual observations, the calculation result is as shown in fig 3. The part where frequency $>$ 0.01 is ground noise, which is actually dominated by white noise, so we fitted the part where frequency $<$ 0.01 and found the slope is about -2.0, so we take $\alpha$=2 when we assume the model. We take $C$=3, the value of $C$ only affects the level of power, and does not affect the solution of the period. We saw that there is an obvious hump in the low frequency part, and its position is consistent with the period we found.

We simulate a large number of (20000) light curves by the Monte Carlo method described in Timmer \& Koenig. 1995. On the one hand, once 20000 light curves were simulated by using even sampling interval, and their LSP was computed. At last, using the spectral distribution of the simulated light curves, local $95\%$, $99\%$, $99.7\%$, and $99.9\%$ confidence contour lines were evaluated, the results are shown in fig 4, the red, blue, green, and purple solid lines represent the confidence level of $95\%$, $99\%$, $99.7\%$, and $99.9\%$, respectively, and the results show that the quasi-periodic oscillation has a high confidence level ($> 5\sigma$ confidence level). On the other hand, we use the WWZ method to analyze these 20,000 light curves and get the power of these 20,000 light curves, using the spectral distribution of the simulated light curves, local $95\%$, $99\%$, $99.7\%$, and $99.9\%$ confidence contour lines were evaluated also. Fig 5 shows the analysis results, the red, blue, green, and purple solid lines represent the confidence level of $95\%$, $99\%$, $99.7\%$, and $99.9\%$, respectively, we can see that the confidence level of the results analyzed by the WWZ method is also higher than $5\sigma$. This method fully randomizes both the amplitude and the phase(e.g., Bhatta. 2018).

\section{Discussion}

In this paper, we have searched quasi-periodic oscillations (QPOs) for BL Lac PKS J2134-0153 in the 15 GHz radio light curve, by means of the LSP and the Weighted Wavelet Z-transform (WWZ) techniques, the period of 15 GHz band is roughly $1743\pm 401$ days by LSP analysis; the period of 15 GHz band by WWZ transform analysis is about $1672\pm 332$ days. It can be concluded that the QPO of BL Lac PKS J2134-0153 is $4.69\pm 0.14$ years ($> 5\sigma$ confidence level). Depending on the long-period light curve, we can predict that the flux of this source will reach its peak again in December 2023.

There are generally three scenarios that are used to explain QPO behavior: 1) the spiral jet scenario; 2) the thermal instability of thin disks scenario; 3) the binary super massive black hole scenario. In the spiral jet scenario, the reason for QPO behavior is due to the relativistic beaming effect, different parts of such a helical jet pass closest to the line of sight, there arises the obvious flux variations because of the change of relativistic beaming effect, even if the emission from the jet have no intrinsic variations. In addition, when the emission blob of the jet moves to us, the viewing angle to the helical motion changes periodically, thereby resulting in QPOs (e.g., Zhou et al. 2018). However, it needs to note that low-frequency radio emission such as 15 GHz is generally considered to be dominated by the extended jet structure of the jet, it is less affected by the beaming effect (e.g., Fan \& Wu. 2018).

In the scenario of thermal instability of thin disks, the thermal instability of a thin disk leads to the periodic cyclic outburst (e.g., Fan et al. 2008). Because there is a certain degree of the link between jet and disk, the observed clumpiness in jet of AGN winds is caused by thermal instability (e.g.,Dannen et al. 2020), and the Accretion-Ejection Instability has been proposed to explain the QPO (e.g., Varni{\`e}re, Rodriguez, \& Tagger. 2002), so the uncertainty of disk causes the related outburst of the jet, which leads to QPO behavior in AGN. Thermal instability of thin disks could produce random light variations in jets with stochastic red noise characteristics, this process is stochastic (e.g., Li et al. 2017). Therefore, this scenario is difficult to be responsible for long-term QPO.

Therefore, we are more promising that the binary supermassive black hole model cause the QPO in the BL Lac PKS J2134-0153, so the BL Lac PKS J2134-0153 is a good binary super massive black hole candidate, in the binary super massive black hole model, the periodic accretion perturbations would be induced by the Keplerian orbital motion of a binary SMBH, or the gravitational torque from a companion engendered the jet-precessional and nutational motions in misaligned disk orbits, thus produce the long-term QPO in blazars (e.g., Katz. 1997; Romero et al. 2000).

If we assume that the binary super massive black hole model leads to the long-term light variations, and we calculate the distance between the primary black hole and the secondary black hole. The mass ratio of binary black holes is in the range of $0.01-0.1$, we obtain 293 black hole masses of the typical blazar (from pieces of literature e.g., Shen et al.2011; Shaw et al. 2012; Liu, Jiang, \& Gu. 2006; Wang, Luo, \& Ho. 2004; Chai, Cao, \& Gu. 2012; Sbarrato et al. 2012; Zhou \& Cao. 2009; Zhang et al. 2012; Xie, Zhou, \& Liang. 2004; Xie et al. 1991), and we calculate the average value of these black hole mass is $\bar{M}=10^{8.6}\,{\rm M_{\odot }}$. We assume that this is the primary black hole mass, and we take a binary black hole mass ratio of 0.1. Therefore, secondary black hole mass can be calculated as $m=\frac{1}{10}\bar{M}=10^{7.6}\,{\rm M_{\odot }}$. We can calculate the relevant parameters of the binary black hole model according to the orbital period.

In this case, we take $P_{obs}=4.69$ years, $\rm M=10^{8.6}\,{\rm M_{\odot }}$, $\rm m=10^{7.6}\,{\rm M_{\odot }}$, according to Kepler's law:
\begin{equation}
\left ( \frac{P_{obs}}{1+z} \right )^{2}=\frac{4\pi ^{2}a^{3}}{G\left ( M+m \right )},
\end{equation} thus we can calculate the distance between the primary black hole and the secondary black hole $a\sim 1.83\times 10^{16} $cm.

\section{Conclusions}

In this paper, we have used the observation data of the OVRO 40-m radio telescope from 2008-01-05 to 2019-05-18 to search for possible quasi-periodic signals in the BL Lac PKS J2134-0153. Our main results are as following:

(1)	Using the Lomb-Scargle periodogram and the weighted wavelet Z-transform techniques, we have found a quasi-periodic signal with a period of $4.69 \pm 0.14$ years ($>$ 5 sigmas confidence level) in the 15 GHz radio light curve of the BL Lac PKS J2134-0153. This is the first time that the quasi-periodic signal has been detected in this source.

(2)	In the scenario where the radio quasi-periodic variability is caused by the precession of the binary super massive black holes, the distance between the primary black hole and the secondary black hole $a\sim1.83\times 10^{16}$ cm.

(3)	Based on the calculation of the quasi-periodic signal of the source, we can predict that the next outburst at the 15 GHz band should reach its peak brightness during December 2023.

This source could be a good binary super massive black hole candidate. We will further monitor the optical variability of the source in the optical band to further verify whether there is the precession of binary super massive black holes.

\section*{Acknowledgements}
N.D are grateful for the support of scientific research fund of Yunnan Provincial Education Department (2021J0715). And this work is supported by the National Nature Science Foundation of China (11663009), and the High-Energy Astrophysics Science and Technology Innovation Team of Yunnan Higher School. This research has made use of data from the OVRO 40-m monitoring program (Richards, J. L. et al. 2011, ApJS, 194, 29) which is supported in part by NASA grants NNX08AW31G, NNX11A043G, and NNX14AQ89G and NSF grants AST-0808050 and AST-1109911. H.Li acknowledges financial supports from the Yunnan Local College Applied Basic Research Projects (2019FH001(-081)), the scientific research fund of talent introduction of Kunming University (YJL2006), and the support of scientific research fund of Yunnan Provincial Education Department (2020J0514).

\section*{Data availability}

The data underlying this article were provided by the OVRO 40-m monitoring program under licence / by permission. Data will be shared on request to the corresponding author with permission of the OVRO 40-m monitoring program.





\bsp	
\label{lastpage}
\end{document}